\documentclass[preprint,epsfig,onecolumn,floats,showpacs]{revtex4}
\usepackage{amssymb}
\usepackage{graphicx}
\usepackage{epsfig}
\usepackage{amsmath}
\usepackage{amsfonts}

\begin{document}

\title{Generation of quantum-dot cluster states with superconducting
transmission line resonator}
\author{Zhi-Rong Lin }
\author{Guo-Ping Guo}
\email{gpguo@ustc.edu.cn}
\author{Tao Tu}
\email{tutao@ustc.edu.cn}
\author{Fei-Yun Zhu}
\author{Guang-Can Guo}

\affiliation{Key Laboratory of Quantum Information, University of
Science and Technology of China, Chinese Academy of Sciences, Hefei
230026, People's Republic of China}
\date{\today }

\begin{abstract}
We propose an efficient method to generate cluster states in spatially
separated double quantum dots with a superconducting transmission line
resonator (TLR). When the detuning between the double-dot qubits transition
frequency and the frequency of the full wave mode in the TLR satisfies some
conditions, an Ising-like operator between arbitrary two separated qubits
can be achieved. Even including the main noise sources, it's shown that the
high fidelity cluster states could be generated in this solid system in just
one step.
\end{abstract}

\maketitle


\textit{Introduction.---} Quantum entanglement is the root in quantum
computation \cite{J. Gruska}, quantum teleportation \cite{Bennett}, quantum
dense coding \cite{Bennett2}, and quantum cryptography \cite{Artur}.
However, it's challenging to create multi-particle entangled states in
experiment. In 2001 Briegel and Raussendorf introduced a highly entangled
states, the cluster states \cite{Hans J. Briegel}, which can be used to
perform universal one way quantum computation. Up to now, various schemes
are proposed to generate cluster states in many different types of physical
systems. Especially, it has been argued that the cluster states can be
generated effectively in solid state system, such as superconductor charge
qubit \cite{Tetsufumi Tanamoto,J.Q. You,Zheng-Yuan Xue} and semiconductor
quantum dot \cite{Massoud,Hui Zhang,Yaakov}.

Electron spins in semiconductor quantum dots are one of the most promising
candidates for a quantum bit, due to their potential of long coherence time
\cite{Taylor,Golovach,Khaetskii}. Producing cluster states in quantum dots,
has been discussed within Heisenberg interaction model \cite{Massoud} and
Ising-like interaction model \cite{Hui Zhang}, where the long-term
interaction inversely ratios to the distance between non-neighboring qubits.
Recently Childress and Taylor \emph{et al.} introduced a technique to
electrically couple electron charge states or spin states associated with
semiconductor double quantum dots to a TLR via capacitor \cite%
{Taylor2,Childress}. The qubit is encoded on the quantum double-dot
triplet and singlet states. The interaction Hamiltonian between the
qubits and the TLR is a standard Jaynes-Cumming (JC) model \cite{Hui
Zhang2}. A switchable long-range interaction can be achieved between
any two spatially separated qubits with the TLR cavity field. This
technique open a new avenue for quantum information implementation.

In this work, we find when the detuning between the qubits transition
frequency and the frequency of the full wave mode in the TLR satisfies some
conditions, an Ising-like operator between arbitrary two separated qubits
can be achieved from the JC interaction. The highly entangled cluster states
can be generated one step with the auxiliary of an oscillating electric
field. Finally, we discuss the feasibility and the cluster states fidelity
of our scheme.

\begin{figure}[tbp]
\includegraphics[width=3in]{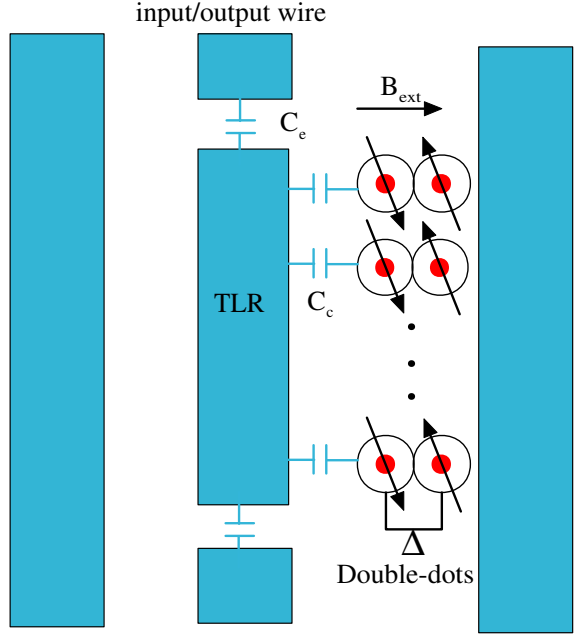}
\caption{(Color online) Schematic diagram of a TLR and several
double quantum dots coupled system. A detailed circuit
representation of the TLR cavity (blue) can be found in Fig.1 of
Ref.\protect\cite{Blais}. The double dots are biased with an
external potential $\Delta$, and capacitive coupling $C_{c}$ with
the TLR. The TLR is connected to the input/output wiring with a
capacitor $C_e$.} \label{Fig1}
\end{figure}

\textit{Preparation of cluster states.---} The system we study
includes N identical double-dot qubits capacitively coupling with a
TLR by $C_{c}$, as shown in Fig. (\ref{Fig1}). The TLR is coupled to
input/output wiring with a capacitor $C_e$ to transmit signals. Two
electrons are localized in double-dot quantum molecule. The two
electrons charge states, spin states and the corresponding
eigenenergies are controlled by several electrostatic gates.

An external magnetic field is applied along axis $z$. At a large static
magnetic field ($100$ mT), the spin aligned states ($\left\vert \left(
1,1\right) T_{+}\right\rangle =\left\vert \uparrow \uparrow \right\rangle $,
$\left\vert \left( 1,1\right) T_{-}\right\rangle =\left\vert \downarrow
\downarrow \right\rangle $) are splitted from spin anti-aligned states ($%
\left\vert \left( 1,1\right) T_{0}\right\rangle =\left( \left\vert \uparrow
\downarrow \right\rangle +\left\vert \downarrow \uparrow \right\rangle
\right) /\sqrt{2}$ , $\left\vert \left( 1,1\right) S\right\rangle =\left(
\left\vert \uparrow \downarrow \right\rangle -\left\vert \downarrow \uparrow
\right\rangle \right) /\sqrt{2}$) due to Zeeman splitting. The notation ($%
n_{l},n_{r}$) indicates $n_{l}$ electrons on the "left" dot and $n_{r}$
electrons on the "right" dot. In addition to the $(1,1)$ subspace, the
doubly-occupied state $\left\vert (0,2)S\right\rangle $ is coupled to $%
\left\vert (1,1)S\right\rangle $ via tunneling $T_{c}$. $\left\vert
(0,2)S\right\rangle $ and $\left\vert (1,1)S\right\rangle $ have a potential
energy difference of $\Delta $. Due to the tunneling between the two
adjacent dots, the $\left\vert (1,1)S\right\rangle $ and $\left\vert
(0,2)S\right\rangle $ hybridize. We can get the double dots eigenstates:
\begin{eqnarray}
\left\vert +\right\rangle &=&-\sin {\theta }\left\vert \left( 1,1\right)
S\right\rangle +\cos {\theta }\left\vert \left( 0,2\right) S\right\rangle ,
\label{EIGENSTATE1} \\
\left\vert -\right\rangle &=&\cos {\theta }\left\vert \left( 1,1\right)
S\right\rangle +\sin {\theta }\left\vert \left( 0,2\right) S\right\rangle ,
\label{EIGENSTATE2}
\end{eqnarray}%
where $\tan {\theta }=-2T_{c}/(\Omega +\Delta )$ and $\Omega =\sqrt{%
4T_{c}^{2}+\Delta ^{2}}$. $\Omega $ is the energy gap between the
eigenstates $\left\vert +\right\rangle $ and $\left\vert -\right\rangle $.
We can choose $\Delta =0$ in order to suppress the fluctuations in control
electronics, then $\left\vert +\right\rangle =(\left\vert \left( 1,1\right)
S\right\rangle +\left\vert (0,2)S\right\rangle )/\sqrt{2}$, $\left\vert
-\right\rangle =(\left\vert \left( 1,1\right) S\right\rangle -\left\vert
(0,2)S\right\rangle )/\sqrt{2}$. The qubit is encoded on the states $%
|+\rangle $ and $|-\rangle $.

An oscillating electric field, which frequency is coincidental with
the qubit energy gap, is applied to the left side gate of the double
quantum dots. The single-qubit operation Hamiltonian in the
interaction picture can be expressed as
\begin{equation}
H_{i}=\eta \left\vert (1,1)S \right\rangle \left\langle
(1,1)S\right\vert=\eta(I+\sigma_{i}^{x})=\eta\sigma_{i}^{x},  \label{Ham2}
\end{equation}
where $\eta=\hat{d}\cdot \hat{E}$, $\hat{d}$ is the dipole of left
quantum dot and $\hat{E}$ is the oscillating electric field.

We consider a TLR of length L, with capacitance per unit length $C_{0}$%
, and characteristic impedance $Z_{0}$. Neglecting the higher energy modes,
we can only consider the full wave mode, with the wavevector $k=\frac{\pi }{L%
}$, and frequency $\omega =\frac{k}{C_{0}Z_{0}}$ \cite{Childress}.
The TLR can be described by the Hamiltonian
$H_{\mathrm{cavity}}=\omega a^{\dagger }a$, where $\hbar =1$,
$a^{\dagger }$, $a$ are the creation and annihilation operators for
the full wave mode of the TLR. In the interaction picture, the
interaction between N qubits and the TLR can be described by the
Hamiltonian \cite{Hui Zhang2}
\begin{equation}
H_{\mathrm{int}}=g_{0}\sum_{i=1}^{N}(e^{i\delta t}a^{\dagger }\sigma
_{i}^{-}+e^{-i\delta t}a\sigma _{i}^{+}),  \label{HJC}
\end{equation}%
where $\sigma _{i}^{+}=\left\vert +\right\rangle \left\langle -\right\vert $%
, $\sigma _{i}^{-}=\left\vert -\right\rangle \left\langle +\right\vert $, $%
\sigma _{i}^{x}=\sigma _{i}^{+}+\sigma _{i}^{-}$, $\delta =\omega -\Omega $.
Here we can presently assume the coupling strength is homogenous. The
overall coupling coefficient can be described by \cite{Childress}
\begin{equation}
g_{0}=\omega \frac{C_{c}}{2C_{tot}}\sqrt{\frac{2z_{0}}{R_{Q}}},
\label{COUPLING}
\end{equation}%
where $R_{Q}=h/e^{2}\approx 26$ k$\Omega $. $C_{tot}$ is the total
capacitance of the double-dot.

If the interaction time $\tau $ satisfies
\begin{equation}
\delta \tau =2k\pi ,  \label{CONDITION1}
\end{equation}%
the evolution operator for the interaction Hamiltonian (\ref{HJC}) can be
expressed as \cite{Zheng}
\begin{equation}
U(\tau )=exp(-i\frac{\lambda }{2}\tau (\sum_{i=1}^{N}{{\sigma _{i}^{x}}}%
)^{2})=exp(-i\lambda \tau \sum_{j>i=1}^{N}\sigma _{i}^{x}\sigma _{j}^{x}),
\label{OPERATOR1}
\end{equation}%
where $\lambda =g_{0}^{2}/2\delta $.

When $\Delta $ is changed to zero, the coupling coefficient between
the qubits and TLR is maximal. An oscillating electric field is
applied to all the qubits at the same time. In the interaction
picture, the total Hamiltonian of the system can be written as
\begin{equation}
H_{tot}=g_{0}\sum_{i=1}^{N}(e^{i\delta t}a^{\dagger }\sigma
_{i}^{-}+e^{-i\delta t}a\sigma _{i}^{+})+\eta \sum_{i=1}^{N}\sigma _{i}^{x}.
\label{TOTOPERATOR}
\end{equation}%
When the operation time $\tau $ satisfies the condition (\ref{CONDITION1}),
we can obtain the total evolution operator
\begin{equation}
U(\tau )=exp(-i\eta \tau \sum_{i=1}^{N}{\sigma _{i}^{x}}-i\lambda \tau
\sum_{j>i=1}^{N}\sigma _{i}^{x}\sigma _{j}^{x}).
\end{equation}%
When $\eta =(N-1)\lambda $, the total evolution operator is given by
\begin{equation}
U(\tau )=exp(-4i\lambda \tau \sum_{j>i=1}^{N}\frac{1+\sigma _{i}^{x}}{2}%
\frac{1+\sigma _{j}^{x}}{2}).  \label{OPERATOR}
\end{equation}

In order to generate the cluster states, the initial state of N
qubits should be prepared in the state
$\bigotimes_{i=1}^{N}\left\vert -\right\rangle
_{i}=\frac{1}{\sqrt{2}}\bigotimes_{i=1}^{N}(\left\vert
0\right\rangle _{i}+\left\vert 1\right\rangle _{i})$, where
$\left\vert 0(1)\right\rangle _{i}=\frac{1}{\sqrt{2}}(\left\vert
-\right\rangle _{i}\pm \left\vert +\right\rangle _{i})$ are the
eigenstates of $\sigma _{i}^{x}$ with the eigenvalues $\pm 1$. Next
we would discuss how to prepare the initial state in experiment.
Firstly we can prepare the two electrons in double quantum dots to
the state $|(0,2)S\rangle $ at a large positive potential energy
difference $\Delta $ \cite{Petta}. Then the $|(0,2)S\rangle
$ can be changed to the state $|-\rangle =(|(1,1)S\rangle -|(0,2)S\rangle )/%
\sqrt{2}$ by rapid adiabatic passage \cite{Taylor}. After the initial state
has been prepared, the qubits would be resonantly coupled with the TLR. We
apply an oscillating electric field to each qubits at the same time. Thus
the initial state would evolve under the total operator (\ref{OPERATOR}). If
the evaluation interaction time satisfies
\begin{equation}
4\lambda \tau =(2n+1)\pi ,  \label{CONDITION2}
\end{equation}%
with $n$ being integer, the spatially separated double quantum dots can be
generated to the cluster states
\begin{equation}
\left\vert \Psi \right\rangle _{N}=\frac{1}{2^{N/2}}\bigotimes_{i=1}^{N}(%
\left\vert 0\right\rangle _{i}(-1)^{N-i}\prod_{j=i+1}^{N}\sigma
_{i}^{x}+\left\vert 1\right\rangle _{i}).  \label{wavefunction}
\end{equation}

The effective coupling coefficient $g_{0}\frac{2T_{c}}{\Omega }$ can be
tuned by external potential $\Delta $. When $\Delta $ is changed, the states
$|0(1)\rangle $ would change according to Eq. (\ref{EIGENSTATE1}), (\ref%
{EIGENSTATE2}), but the expression (\ref{wavefunction}) of the cluster
states is unchanged. When the cluster states is generated at the time of $%
\tau $, we can remove the oscillating electric field and change $\Delta $ to
discouple all the qubits to the TLR. Then the cluster states can be
preserved.

\textit{Feasibility of the scheme.---} The sample of the TLR and
quantum dots coupled system can be obtained in a two-step
fabrication process on a GaAs/AlGaAs heterostructure. Firstly,
quantum dots are formed in the two-dimensional electron gas below
the surface, using electron beam lithography and Cr-Au
metallization. Then the TLR is fabricated by conventional optical
lithography \cite{Wallraff}. The main technical challenges for
experimental implementation of our proposal are the design and
nanofabrication of the sample \cite{Schusterthesis}. The diameter of
the quantum dot is about $400$ nm, and the corresponding capacitance
of the double-dot $C_{tot}$ is about $200$ aF. The distance between
the two double-dot molecules should be $4$ $\mu $m which is tenfold
of the distance between two quantum dots within a double-dot. Thus
we can neglect the interaction between the double-dot molecules
safely \cite{Guo}. Since the energy gap between $\left\vert
+\right\rangle $ and $\left\vert -\right\rangle $ is about $10$ $\mu
$eV at the operation point, the experimental manipulation should be
implemented in dilution refrigerator
with temperature below $100$ mK. Both the conditions (\ref{CONDITION1}) and (%
\ref{CONDITION2}) are satisfied whenever $\delta =g_{0}\sqrt{4k/(2n+1)}$.
From Eq. (\ref{COUPLING}) the coupling coefficient $g_{0}$ can be up to $%
\omega /16$ with a large coupling capacitor $C_{c}\approx 2C_{tot}\approx
400 $ aF. For $k=1$, $n=0$, $\omega /2\pi =2$ GHz and $g_{0}/2\pi =125$ MHz,
the operation time of the generation of cluster states is $\tau =\frac{2\pi
}{\delta }=4$ ns.

\textit{Decoherence.---} In our system, the main decoherence processes are
the dissipation of the TLR, the spin dephasing, charge relaxation and
additional dephasing of the double-dot molecules. The dissipation of the TLR
occurred through coupling to the external leads can be described by the
photon decay rate $\kappa =\omega /Q$, where $Q$ is the quality factor. For $%
Q=1\times 10^{5}$, $\omega =2\pi \times 2$ GHz in our situation, the
photon decay time $1/\kappa \approx 50$ $\mu $s is $4$ orders longer
than the operation time $\tau $. Therefore the cavity loss can be
neglected in our situation.

Nuclear spins are one of the main noise sources in semiconductor quantum
dots via hyperfine interaction. The hyperfine field can be treated as a
static quantity, because the evolution of the random hyperfine field is
several orders slower ($>10$ $\mu $s) than the electron spin dephasing. In
the operating point, the most important decoherence due to hyperfine field
is the spin dephasing between the states $|(1,1)S\rangle $ and $%
|(1,1)T_{0}\rangle $. By suppressing nuclear spin fluctuations \cite%
{Marcus2008}, the spin dephasing time obtained by quasi-static
approximation can be $T^{*}_{2}=\hbar /(g\mu _{B}\langle \Delta
B_{n}^{z}\rangle _{rms})\approx 1$ $\mu $s, where $\Delta B_{n}^{z}$
is the nuclear hyperfine gradient field between two dots and rms
denotes a root-mean-square time-ensemble average. The coupling to
the phonon bath will lead to the relaxation of the charge freedom.
Using the spin-boson model, the relaxation time of the qubits can be
obtained by Fermi-Golden rule \cite{Taylor2}. The charge relaxation
time $T_{1}$ is about $1$ $\mu $s at the operation point. Additional
dephasing is assumed to arise from the low frequency fluctuations of
the control electronics, which typically have the $1/f$ spectrum. In
our system, it's assumed that the origin of $1/f$ noise is the
random drift of the gate bias when an electron tunnel in or out of
the metallic electrode. Assuming $1/f$ noise subject to Gaussian
statistics, we found the addition dephasing time $T_{2,\alpha }\sim
\Omega T_{2,bare}^{2}$ is about $100$ ns
at the optimal working point ($\Delta \approx 0$), where $%
T_{2,bare}$ will be discussed below in detail and can be up to $10$ ns. Thus
the total operation time of the present proposal $\tau \approx 4$ ns is much
shorter than the spin dephasing time, charge relaxation and additional
dephasing time of the qubits.

\textit{Fidelity of the cluster states.---} For simplicity, we assume the
control electronics fluctuations are Gaussian. These noises would lead to
the fluctuations of the parameter $\lambda $ via the electric potential
difference $\Delta $. Suppose $\Delta _{i}(t)=\Delta +\epsilon _{i}(t)$, $%
\left\langle \epsilon _{i}(t)\right\rangle =0$, $\left\langle \epsilon
_{i}(t)\epsilon _{j}(t{^{\prime }})\right\rangle =\int S_{ij}(\omega
)e^{i\omega (t-t^{\prime })}d\omega $ ($i$ labeling the $i$-th qubit). The
fluctuations of the oscillating electric field would result in the
fluctuations of the parameter $\eta $. The fluctuations of $\lambda $ and $%
\eta $ would add an unwanted phase $\theta _{i}$ to the desired value $\pi $
\cite{TAME}. Including the fluctuations, the evolution operator (\ref%
{OPERATOR}) should be rewritten in the form of
\begin{widetext}
\begin{equation}
U(\tau)=exp(-i\pi\sum_{j>i=1}^{N}\frac{1+\sigma_{i}^{x}}{2}\frac{1+\sigma_{j}^{x}}{2})
exp(-i\int_{0}^{\tau}\delta\eta(t)dt
\sum_{i=1}^{N}{\sigma_{i}^{x}})exp(-i\int_{0}^{\tau}\delta\lambda(t)dt
\sum_{j>i=1}^{N}{\sigma_i^{x}\sigma_j^{x}}),
\end{equation}
\end{widetext}where $\lambda (t)=\lambda +\delta \lambda (t)$ and $\eta
(t)=\eta +\delta \eta (t)$. Since $\delta \lambda +\frac{\delta \eta }{N-1}$%
, $\delta \eta +(N-1)\delta \lambda $ are larger than $\delta \lambda $, $%
\delta \eta $, we can write the unwanted phase $\theta _{i}=\int_{0}^{\tau }{%
4(\delta \lambda (t)+\frac{\delta \eta (t)}{N-1})dt}=\theta _{1,i}+\theta
_{2,i}$, where $\theta _{1,i}=\int_{0}^{\tau }{4\delta \lambda (t)}dt$ and $%
\theta _{2,i}=\int_{0}^{\tau }{4\frac{\delta \eta (t)}{N-1}}dt$.

Since $\delta$, $\lambda$ satisfy Gaussian distribution, $\theta_{1,i}$, $%
\theta_{2,i}$, $\theta_i$ have Gaussian distribution $G(0,\sigma_{1,i}^{2})$%
, $G(0,\sigma_{2,i}^{2})$, $G(0,\sigma_i^{2})$. Ignoring the correlative
fluctuations, the variance of $\theta_{1,i}$ at the optimal working point is
\begin{align}
\sigma_{1,i}^{2} =(\frac{2g_0^{2}}{\Omega\delta^{2}})^{2}\left\langle
(\int_{0}^{\tau} \epsilon^{2}(t) dt)^{2} \right\rangle,
\end{align}
where
\begin{equation}
\begin{array}{llll}
\left\langle (\int_{0}^{\tau}{\ \epsilon^{2}(t)
dt})^{2}\right\rangle & = &
(\int S_{i}(\omega)d\omega)^{2}\tau^{2} &  \\
&  & +2(\int S_{i}(\omega)\frac{\sin\omega\tau}{\omega\tau}%
d\omega)^{2}\tau^{2}. &
\end{array}%
\end{equation}

For the low frequency noise, $S_{i}(\omega )$ has a high frequency cutoff $%
\gamma \ll \frac{1}{\tau }$. Therefore we can get $\left\langle
(\int_{0}^{\tau }{\epsilon ^{2}(t)dt})^{2}\right\rangle =3(\int S_{i}(\omega
)d\omega )^{2}\tau ^{2}$. Assuming $\frac{1}{T_{2,bare}^{2}}=\int
S_{i}(\omega )d\omega $, we can obtain the variance $\sigma _{1,i}^{2}=\frac{%
12g_{0}^{4}}{\delta ^{4}}(\frac{\tau }{\Omega T_{2,bare}^{2}})^{2}$. Taking $%
T_{2,bare}\approx 10$ ns from the Ref. \cite{Petta,Bracher,Koppens},
the variance of $\theta _{1,i}$ is $\sigma _{1,i}=0.022\pi $. The
fluctuations of the oscillating electric field root in the
electronics noise. The fluctuations  can be reduced in a small value
with better high- and low-frequency filtering technique. Supposing
$\sigma _{\Delta \eta }/\Delta \eta \approx
2\%$, the variance of $\theta _{2,i}$ is $\sigma _{2,i}=0.006\pi $. So $%
\theta _{i}$ has an Gaussian distribution $G(0,(0.023\pi )^{2})$. The
fidelity of N qubits cluster states is calculated according to the formula $%
F=|2^{-N}\sum_{z_{i}}\prod_{j=1}^{N-1}(\int \frac{1}{\sqrt{2\pi }\sigma _{i}}%
e^{-\frac{\theta _{j}^{2}}{2\sigma _{i}^{2}}}e^{i\theta _{j}}d\theta
_{j})^{z_{j}z_{j+1}}|^{2}$ from Ref. \cite{TAME}, as shown in Fig. (\ref%
{Fig2}). The fidelity of a 30-qubit cluster states can be $96.2\%$.

\begin{figure}[tbp]
\includegraphics[width=3in]{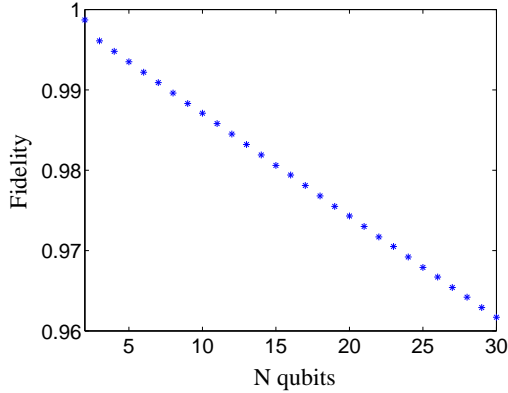}
\caption{The fidelity of N-qubit cluster states.}
\label{Fig2}
\end{figure}

\textit{Conclusion and Discussions.---} Distinguished from cavity
quantum electrodynamics in atomic quantum information processing,
our scheme can realize the long-range interaction among the
double-dot molecules with the TLR in a solid micro-chip device. This
technique can couple the static qubit in the solid state system to
the flying qubit (the cavity photon) \cite{Taylor2}. Compared with
other schemes, the present proposal based on quantum dot molecules
has four potential advantages: integration and scaling in a chip,
easy-addressing, high controllability, and long coherence time
associated with the electron spin. As discussed above, the
preparation of the initial state can be easily implemented without
inter-qubit coupling in our scheme. When the initial state has been
prepared, the quantum-dot cluster states can be produced with only
one step. The cluster states can be preserved easily by switching
off the coupling between the qubits and TLR cavity field.

In conclusion, we proposed a realizable scheme to generate cluster states
only one step in a new scalable solid state system, where the spatially
separated semiconductor double-dot molecules are capacitive coupling with a
TLR. An effective, switchable long-range interaction can be achieved between
any two double-dot qubits with the assistance of TLR cavity field. The
experimental related parameters and the possible fidelity of generated
cluster states have been analyzed. Due to the long relaxation and dephasing
time at the optimal working point, the present scheme seems implementable
within today techniques.

This work was funded by National Basic Research Programme of China
(Grants No. 2009CB929600, No. 2006CB921900), the Innovation funds
from Chinese Academy of Sciences, and National Natural Science
Foundation of China (Grants No. 10604052, No.10804104, No.
10874163).


\end{document}